\documentstyle[11pt,newpasp,twoside,epsf]{article}
\def\edcomment#1{\iffalse\marginpar{\raggedright\sl#1\/}\else\relax\fi}
\reversemarginpar

\begin{document}
\title{Faint and peculiar objects in GAIA: results from GSC-II}
\author{D. Carollo, A. Spagna, M.G. Lattanzi, R.L. Smart}
\affil{INAF- Osservatorio Astronomico di Torino, I-10025, Torino,
Italy}
\author{S.T. Hodgkin}
\affil{Cambridge Astronomical Survey Unit, Institute of
    Astronomy, Cambridge, CB3 0HA, UK}
\author{B. McLean}
\affil{Space Telescope Science
    Institute, MD 21218, USA}
\begin{abstract}

The one billion objects in the GSC-II make up a formidable data
set for the hunt of peculiar and rare targets such as late type
stars, white dwarfs, carbon dwarfs, asteroids, variable stars,
etc. Here we present a survey  to search for ancient cool white
dwarfs, which led to the discovery of several stars with peculiar
spectral distributions and extreme physical properties.  Finally,
we discuss the impact of the GAIA mission with respect to these
peculiar and faint white dwarfs.

\end{abstract}

%\section{Introduction}

\section{The Second Guide Star Catalogue}

The Second Guide Star Catalogue (GSC-II) project (Lasker et al.
1995, McLean et al. 2000) is a collaborative effort between the
Space Telescope Science Institute (STScI) and the Osservatorio
Astronomico di Torino (OATo) with the support of the European
Space Agency (ESA) - Astrophysics Division, the European Southern
Observatory (ESO) and GEMINI. The aim of this project is the
construction of an all-sky catalogue containing classifications,
colors, magnitudes, positions and proper motions of $\sim$ 1
billion objects down to the magnitude limit of the plates ($B_{J}$
$\sim$ 22.5). At the moment, GSC-II is one of the largest stellar
catalogue with the only comparable one being USNO-B (Monet et al. 2002).

GSC-II is based on about 7000 photographic Schmidt plates (POSS
and AAO) with a large field of view (6.4$^\circ \times
6.4^\circ$). All plates were digitized at STScI utilizing modified
PDS-type scanning machines with 25 $\mu$m square pixels (1.7
$''$/pixel) for the first epoch plates, and 15 $\mu$m pixels (1
$''$/pixel) for the second epoch plates. Each digital copy of the
plate was analyzed by means of a standard software pipeline which
performs object detection and computes parameters, features and
classification for each identified object. Position and magnitude
for each object was found from astrometric and photometric
calibrations which utilized the Tycho2 (H{\o}g et al.\ 2000) and
the GSPC-2 (Bucciarelli et al.\ 2001) as reference catalogs.\\ The
first public release of GSC-II (GSC2.2) was delivered in June 2001
and contains 445,851,237 objects
%\footnote {Data are available on-line at CDS in Strasburg or at
%\tt http://www-gsss.stsci.edu/ gsc/GSChome.htm.}
 down to $B_{J} < 19.5$ and $R_{F}
< 18.5$ providing positions with an average accuracy of 0.2
arcsec, photographic photometry $B_{J}$ and $R_{F}$ with 0.15-0.2 mag
accuracy and classification (stellar/extended objects) accurate to 90$\%$.

\section{Search for nearby Halo White Dwarfs}

Cool white dwarf (WD) stars are the remnants of stars which were
born when the Milky Way was very young. A WD cools and fades in a
well defined manner, thus the WD luminosity function is imprinted
with the star formation history of the Galaxy back to its very
beginning. In particular, detecting the end of the WD sequence
will provide a direct measure of the age of the Galaxy and its
fundamental components, i.e.\ the disk, thick disk and the halo
(Fontaine, Brassard \& Bergeron 2001). The usefulness of WDs as
stellar chronometers has been stimulated by the recent progress in
the cooling models based on non-grey atmospheres and refinements
of the internal physics (e.g. Hansen 1999, Chabrier et al.\ 2000)
for both DA and non-DA WDs with $T_{\rm eff}< 4000$ K, which are
the temperatures expected for ancient halo WDs. In this range,
cooling WDs with hydrogen atmosphere start to become fainter but
bluer because of the strong H$_2$ opacity due to the collision
induced absorption (CIA) towards longer wavelengths, whereas
helium atmosphere WDs continue to redden.

In addition, it has been suggested that Pop.II WDs could
contribute significantly to the baryonic fraction of the dark
Halo. In fact they are obvious candidates for the MACHOs revealed
by the LMC microlensing surveys (Alcock 2000) which seem to
indicate that $\sim$ 20\% of the dark matter is tied up in objects
with $\sim$ 0.5 M$_ \odot$. The most extensive survey to date
(Oppenheimer et al.\ 2001) provides a lower limit on the space
density of $\rho\sim 10^{-4}$ pc$^{-3}$, that is, 5 times larger
than expected from the canonical stellar halo, and $\sim 1$\% of
the expected local dark halo density. These results are still a
matter of debate. In fact Reid, Sahu and Hawley (2001) claimed
that the kinematics of the Oppenheimer sample is consistent with
the high-velocity tail of the thick disk. Moreover these stars
have a spread in age that is more consistent with a thick disk
population (Hansen 2001).

The aim of our survey is to search for halo WDs using plate
material from the GSC-II in the Northern hemisphere and improve
the measurements of halo WDs space density. Also, we will confront
the WD models with our sample of cool ancient objects in order to
improve the cooling tracks of WDs with $T_{\rm eff}<4000$ K.
Although WDs should be a typical component of the Halo, such
objects are very difficult to observe because they are extremely
faint. In fact, theoretical cooling tracks by Chabrier et al.\
(2000) predict an absolute magnitude of $M_V=16.2$ and 17.3 for a
0.6 M$_\odot$ WD of 10 and 13 Gyr respectively (excluding the
nuclear burning phases). Objects with these magnitudes are
observable only within a few tens of parsecs with the GSC-II
material which contains objects down to the plate limits of 22.5,
20.8 and 19.5 mag for the blue $B_J$, red $R_F$ and infrared $I_N$
plates respectively.

\begin{figure}[t]
%\plottwo{rpm.eps}{spectra1.eps}
%{12cm}{0}{45}{45}{-130}{100}
\plotfiddle{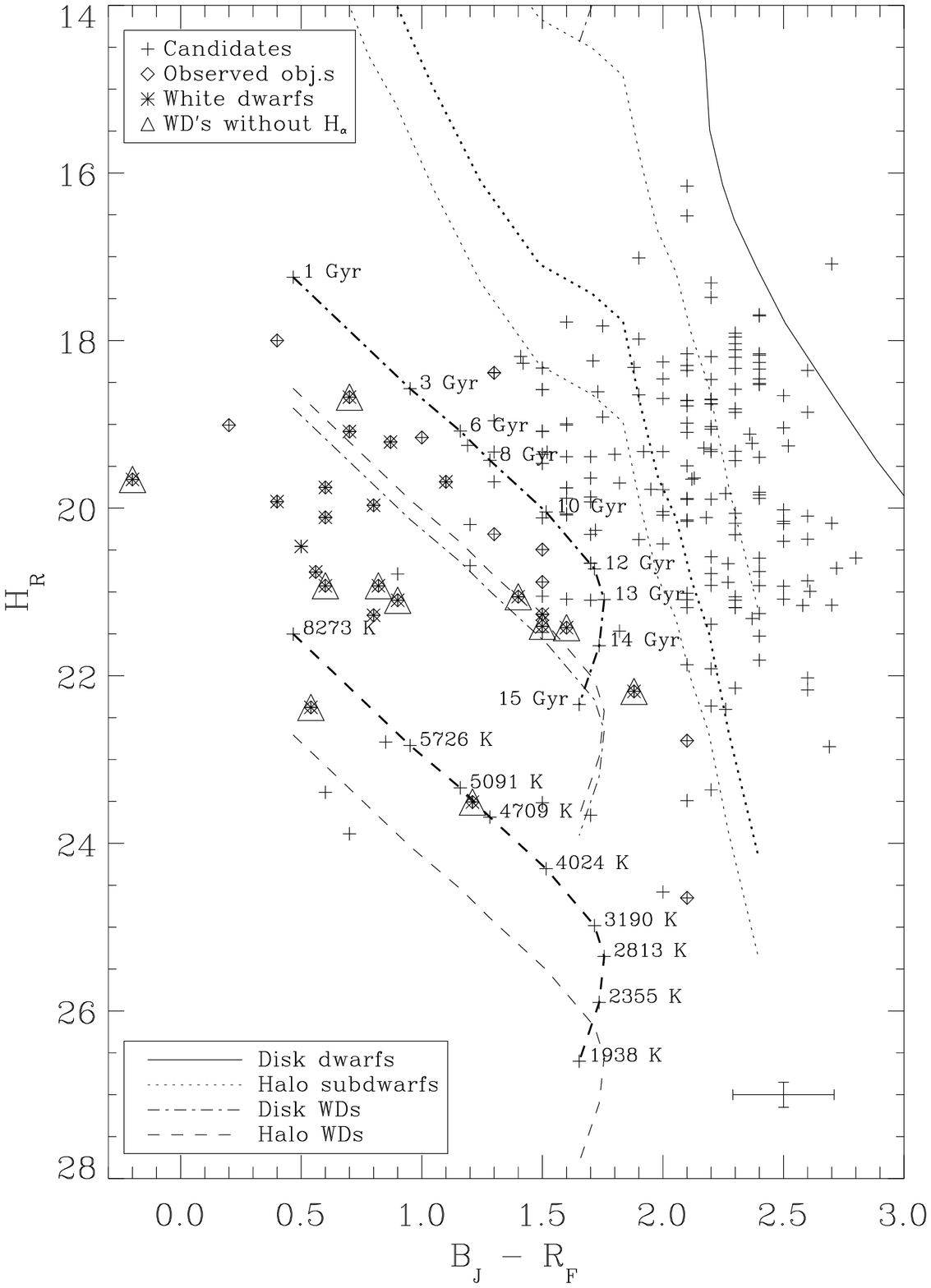}{7cm}{0}{50}{50}{-240}{-120}
\plotfiddle{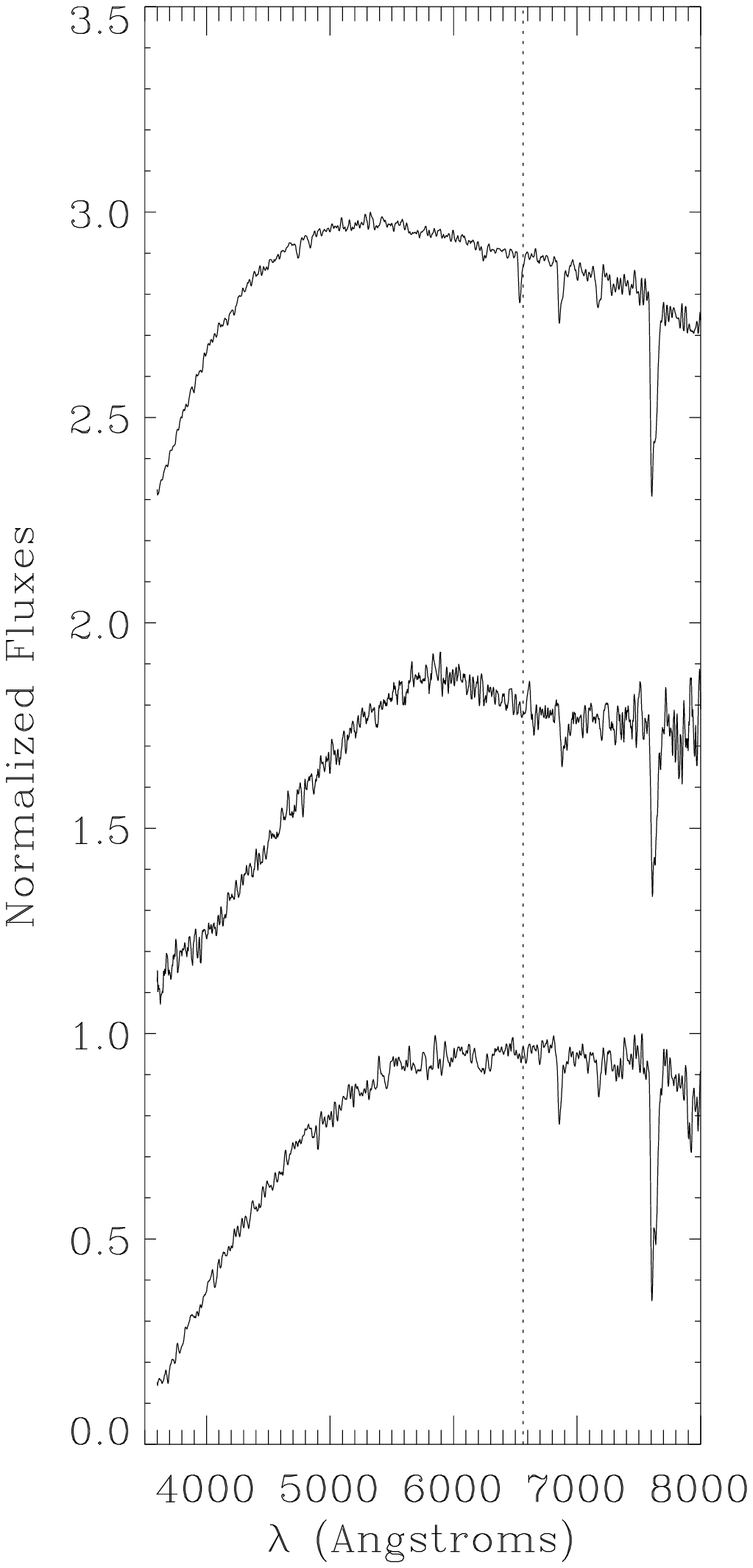}{3cm}{0}{53}{50}{10}{-10}

\caption{Left panel: the RPM diagram for an area of 800 square
degrees (see text). Right
 panel: spectra of new cool WDs. Dotted line indicates H$_{\alpha}$.}
\end{figure}

\subsection{Plate material, processing and selection criteria}

Our survey covers an area of $\sim$ 1300 square degrees which
corresponds to 40 regions in the sky, mostly located toward the
North Galactic Pole (NGP). In order to detect high proper motion
objects, we processed POSS-II plates (blue, red, infrared) with
epoch difference $\Delta$t $\sim$ 2-10 yr by means of the standard
GSC-II pipeline. Also, we performed object matching and derived
proper motions using the procedure described in Spagna et al.\
(1996), then faint ($R_{F}>16$ mag) and fast moving
($0.3<\mu<2.5''$/yr) stars were identified. Each target was
checked by a visual inspection of POSS-I and POSS-II plates in
order to reject the false detections (e.g.\ mismatches and
binaries) and to confirm its proper motion. Another very useful
parameter for the selection of the targets is the reduced proper
motion (RPM),  $H = m + 5 \log\mu -5$. The RPM diagram, $H_{R}$
vs. ($B_{J}-R_{F}$), was adopted to identify faint objects with
high proper motion and to separate disk ad halo WDs from late type
dwarfs and subdwarfs. Figure 1 (left panel) shows the RPM diagram
for a set of regions. Here, the thick solid and dotted lines show
the locus of the disk dwarfs and the halo subdwarfs based on the
10 Gyr isochrones down to 0.08 M$_\odot$ from Baraffe et al.
(1997, 1998) with [Fe/H]=0 and -1.5, respectively. Dashed and
dot-dashed lines show the cooling tracks of 0.6 M$_\odot$ WD's
with hydrogen atmosphere from Chabrier et al (2000). We adopted
mean tangential velocities (towards the NGP) of $V_{T}=38$ km/s
(disk) and 270 km/s (halo). Thin lines indicates the 2$\sigma$
kinematics thresholds. Finally, spectral analysis is required for
a confirmation of the nature of the selected candidates.

\begin{figure}[t]
\plotfiddle{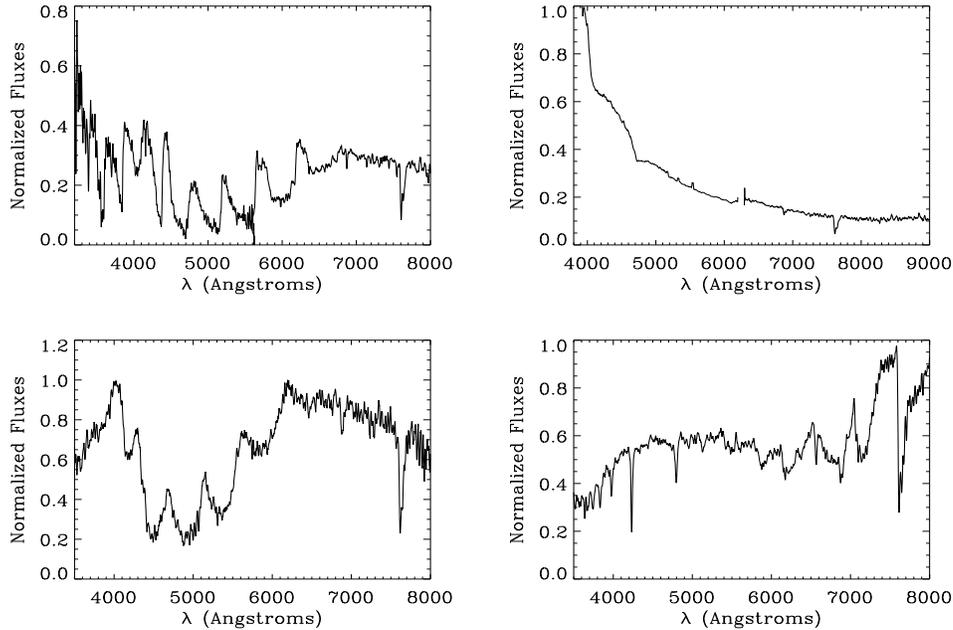}{8cm}{90}{55}{50}{210}{0} \caption{A
sample of peculiar objects. Top left: the peculiar DQ WD with
strong C$_2$ Deslandres-d' Anzabuja and Swan bands. Top right: a
very hot magnetic WD candidate. Bottom left: a magnetic DQ WD.
Bottom right: a binary system (WD+dM)}
\end{figure}

\subsection{Spectroscopic follow-up and preliminary results}

Low resolution spectroscopy is suitable to recognize the spectral
type and the main chemical composition of the stars. Spectroscopic
observations were carried out with the 3.5 m TNG (La Palma), the
4.2 m WHT (La Palma), and the 3.5 m at Apache Point Observatory
(USA). Most of the targets were observed in the first semester
2002 at TNG using the low resolution spectrograph DOLORES (Device
Optimized for Low Resolution) with the LR-B Grism1 which gave a
nominal dispersion of 2.8 \AA/px and useful wavelength coverage
from 3000 to 8800 \AA. We performed spectroscopic follow-up for
candidates from 800 square degrees (1/50 of the sky) which
corresponds to $\sim 60\%$ of our total area. The number of halo
WD candidates after the selection criteria was 47 and we obtained
32 spectra plus JHK infrared photometry  for 12 stars during 3
nights. The results are remarkable: of the 32 observed targets, 23
are WDs and 12 have no H$\alpha$ line. We also found 4 M dwarfs, 2
subdwarfs, a binary system (dM+WD) and 3 interesting peculiar
objects. The left panel of Figure 1 shows the RPM diagram for
these 800 square degrees. The observed objects and those
classified as white dwarfs are marked with different symbols. The
right panel of Figure 1 shows a few confirmed cool WDs in our
sample, including a ''coolish'' DA (top spectrum) with a weak
H$\alpha$ line and two cool WDs, while Figure 2 shows the peculiar
objects.

\section{Peculiar objects and classification problems}

An unexpected result of this survey is the discovery of a
significant fraction of objects with a very complex nature. Some
examples are presented in Figure 2, where the top left panel shows
a peculiar DQ WD, with extremely strong $C_{2}$ absorption bands,
while the bottom left shows a magnetic carbon rich WD. On the top
right is a probable very hot magnetic WD and the bottom right an
unresolved binary system WD+dM. We point out that all these cases
could not be classified properly till their spectra became
available. Even when spectra are available, the classification can
be tricky for objects with extreme physical properties and no
previous observations or good theoretical models. This was the
case of the peculiar carbon rich WD named GSC2U J131147.2+292348
(Fig. 1, top left). The object is fast moving ($\mu\simeq 0.48$
arcsec~yr$^{-1}$), and faint ($V\simeq 18.7$). A check on the
SIMBAD database revealed that the star was not in the NLTT
catalogue (Luyten 1979) but, quite surprisingly, was listed as a
quasar candidate (object OMHR 58793) by Moreau \& Reboul (1995),
who measured an UV excess but did not detect any proper motion,
perhaps because of a cross-matching error. The real nature of this
object was realized after a thorough analysis of its spectrum with
the support of infrared photometry (Carollo et al.\ 2002), even
though the lack of adequate models in the literature was a serious
problem.

\section{The impact of the GAIA mission}
The main impact of GAIA  with respect to the faint and nearby
objects, such as the large variety of WD types, will be the
determination of accurate distances by means of the {\it
trigonometric parallaxes}  for {\it all} the objects detectable in
the solar neighborhood down to $V\approx 20$ mag. Distances will
directly provide the absolute magnitudes which will permit a
robust, even if preliminary, identification of these objects as
WDs.  Thus, the fact that the information from the GAIA
spectrograph and the medium band photometric system will not be
available for the faintest objects is not as dramatic a problem as
in the case of current ground based surveys.

We expect that GAIA will carry out a complete {\it census} of WDs,
including the faintest ancient and cool WDs previously discussed.
To this regard, GAIA will provide a complete and unbiased sample
(i.e.\ not kinematically selected).  Moreover, accurate tangential
velocities will be derived from proper motions also in the cases
where radial velocities are not available, and will help to
separate the halo and disk WDs. Of course, a large fraction of
non-DA WDs including a certain percentage of peculiar objects will
also be detected by GAIA. Broad band photometry should help to
identify these cases by means of  their {\it anomalous} colors
also for the dimmest objects without any further
spectro-photometric data.
 Clearly, this is a challenging and non trivial issue for
 the classification task of the GAIA data reduction.
 Moreover, among the many science cases that GAIA data will
address, these objects point out the logical necessity to
complement the astrometric and spectro-photometric observations of
peculiar or unclassified objects with a spectroscopic
follow-up  with large ground based telescopes.
 Fortunately, it will probably not be difficult to obtain
observing time with 4-8 meters telescopes at the epoch when the
first GAIA results will be delivered ($\sim$ 2015?).

Finally, we mention the fact that objects with such a peculiar
spectral distribution as those shown in Figure 2 could be affected
by residual systematic errors due to a chromaticity effect (Gai et
al. 1998, Lindegreen 1998) not properly corrected by the standard
astrometric calibrations. However, this does not seem a critical
problem for nearby objects (i.e.\ having large values of parallax
and proper motion). Also, the possibility that the multiple
observations of these high proper motion objects are not correctly
matched in crowded regions can probably be avoided by means of a
robust matching algorithm.

\end{document}